\documentstyle[11pt,magfields_pasp,twoside,epsf]{article}
\def\edcomment#1{\iffalse\marginpar{\raggedright\sl#1\/}\else\relax\fi}
\reversemarginpar

\def\kms{\hbox{km$\,$s$^{-1}$}}
\def\Halpha{\hbox{H\,$\alpha$}}
\def\etal{\hbox{et al.}}

\def\SiXIII{\hbox{Si\,{\sc xiii}}}
\def\SXV{\hbox{S\,{\sc xv}}}

\def\vinf{\hbox{$v_{\infty}$}}
\def\zpup{\hbox{$\zeta$ Pup}}
\def\zori{\hbox{$\zeta$ Ori}}
\def\bcma{\hbox{$\beta$ CMa}}

\def\ecma{\hbox{$\epsilon $ CMa}}

\def\Msunyr{\hbox{$M_\odot\,$yr$^{-1}$}}

\def\1half{\hbox{\tiny 1/2}}
\def\3half{\hbox{\tiny 3/2}}

\begin{document}
\title{Observable Effects of B fields on the Winds and Envelopes of Hot Stars}
\author{Joseph P. Cassinelli}
\affil{Astronomy Dept. Univ. of Wisconsin, 475 N. Charter St. 
Madison WI. 53706, USA}

\begin{abstract}

Magnetic fields on hot stars can produce a variety of interesting effects on
the velocity, density, and temperature structure in the winds from the
stars. The fields can produce a longitudinal dependence of the mass flux,
which would lead to cyclical variability such as that seen in optical and UV
spectra of many early-type stars. The fields can channel and spin-up the
outflow, which appears to be needed to explain the disk-like density
enhancements around Ap and Be stars. Magnetic confinement of hot gases can
occur and be responsible for the anomalously high EUV fluxes seen in B
giants, and the anomalous high ion X-ray line emission that is seen in
recent CHANDRA observations. Special attention is given to the reports at
this meeting that A stars develop magnetic Ap phenomenon only after about 30
percent of their main sequence lifetimes. A similar delay occurs for the
emission line Be stars. It is suggested that these delays are related to the
time it takes for fields to rise through the sub-adiabatic envelope to the
surface starting from the interface between the convective-core and
radiative-envelope where they are generated.
\end{abstract}

\section{Introduction}

Early type stars have several major differences from the Sun and other cool
stars, and these lead to a new array of magnetic phenomena. The early type
stars tend to be very luminous, and the emission line Of stars and the OB
supergiants lie rather near the Eddington upper luminosity limit. Because of
the strong radiation fields, the most important properties of their winds
can be explained by line driven wind theory as first developed by Castor,
Abbott and Klein (CAK, 1975). With the modifications of CAK theory developed
by Friend \& Abbott (1986) and Pauldrach, Puls, \& Kudritzki (1986), we can
now understand why the mass loss rates of luminous early-type stars are in
the range of $10^{-6}$ to $10^{-5}$ \Msunyr, and why the terminal
velocities, \vinf, are generally in the range of 1500 to 3000 \kms. However,
there are some properties of hot stars that are not easily explained by line
driven wind theory, and magnetic fields appear to play a role in those. For
example, in line driven wind theory, one should expect that the radiation
field should be axially symmetric. There could be a latitude dependence
owing to polar brightening in rapidly rotating stars, however as we have
heard in the paper by Henrichs, there are dependences of the winds on
longitude, and this leads to the cyclic variability in UV wind diagnostics.
The O, B, and A stars also tend to be rapid rotators. In the case of the
emission line Be stars the rotation speeds are on average about 70 percent
of the critical rotation speed at which the surface is rotating at
Keplerian speed. These stars have equatorial disks, and a consensus has
recently been reached that the disks are Keplerian (Smith, Henrichs, \&
Fabregat, 2000). So the major problem for understanding the Be stars is
explaining how a star rotating at sub-Keplerian speed can produce a
Keplerian disk. Co-rotating magnetic fields could lead to the needed
transfer of mass and angular momentum from the star to the disk.

\section{Evidence Across the Spectrum}

Here I will briefly discuss the evidence for magnetic fields that are seen
in early-type stars at wavelengths ranging from X-ray to the optical. The
X-rays from early type stars are usually explained as arising from shocks
that are distributed throughout the relatively dense winds. The recent
launches of the Chandra and XMM satellites have led to a great improvement
of our understanding of the source of the X-rays because of the high
spectral resolution they provide. The X-rays in the winds are thought to be
produced by instabilities in line driven winds. However, models for shock
formation such as those by Lucy (1982) and Feldmeier et al (1997) and others
have consistently failed to predict the high level of X-ray emission in the
brightest stars such as \zpup\ (O4f) (Cassinelli et al 2001, Kahn et al;
2001) and \zori\ (O9.5 Ia) (Cassinelli and Swank, 1983, Waldron \&
Cassinelli, 2001). The X-ray spectral lines allow us for the first time to
determine the radius at which the X-rays are originating. In the case of the
two stars mentioned, there is evidence that X-rays from the highest ion
stages are originating so close to the star that the local wind and its
maximal strength shocks could not produce the high ion stages such as
\SiXIII\ and \SXV\ that are observed. Waldron \& Cassinelli (2001) and
Schulz et al (2000) in particular, have suggested that magnetic fields are
largely responsible for the very hot gases. In the case of lower luminosity
near-main sequence B stars, Cohen et al (1997) found that the X-ray
luminosities are larger than could be generated by the emission measure of
the entire wind. This also suggests that there could be an extra source of
X-ray emitting material that is perhaps magnetically confined.

There is also indirect evidence for magnetic fields in stars observable in
the extreme ultraviolet. Only two early-type stars \ecma\ (B2 II) and 
\bcma\ (B1 II-III) could be observed with the EUVE satellite.
These two stars lie in a rarefied tunnel in the ISM and the attenuation of
their EUV light is unusually small. For \ecma, Cassinelli \etal\ (1995)
found the star to have a photospheric radiation flux level in the wavelength
range 500 to 700 \AA\ that exceeds the values predicted by model atmospheres
by a factor of about 30. This excess could be produced if the star were to
have an outer atmospheric zone like a chromosphere, that is hotter than
expected from radiative equilibrium model predictions. Such a zone is
present in the solar atmosphere, but is not otherwise observed in O and B
stars; nor is one expected since hot stars do not have an outer convection
zone. Thus, some sort of mechanical heating seems to be required. A possible
explanation is provided by observations of the other EUVE stellar source,
\bcma. This star shows an EUV continuum flux that exceeds model predictions
by about a factor of 5 (Cassinelli \etal\ 1996). It is a bright pulsating
variable of the $\beta$ Cephei class, and the star $\beta$ Cephei itself has
been measured via the Zeeman effect methods to be an oblique rotator with a
field of about 100 Gauss (Henrichs \etal\ 2001, and at this meeting). The
connection between the heating that is needed on one bright member of this
class of variables, and the magnetic field that is seen on another is not
yet clear, but the possibility that a magnetic field is also present on
\bcma\ is plausible and very interesting. The results might mean that
heated outer atmospheres are common among B stars.

In the ultraviolet, there are very many observations of hot star spectra.
Perhaps of most interest here are the observations of cyclic behavior of the
discrete absorption components (DAC's) of moderately strong UV resonance
line profiles (Prinja \etal\ 1992; Kaper \etal\ 1996). The progression of
the DAC features across the line profiles has been explained by the
co-rotating interaction region (CIR) model of Cranmer and Owocki (1996).
CIR's are present in the solar wind, and Mullan (1984) had proposed that
they are present also in hot star winds. What is needed in the model, are
sectors of fast stellar wind streams adjacent to sectors of slower moving
material. This is not expected to occur in purely radiation driven winds.
However, it is well known in solar work that magnetic fields can lead to
fast stream - slow stream outflow. MacGregor (1988) investigated a way in
which this could occur on a star with a line driven wind by magnetic
channeling of the flow, and Cassinelli and Miller (1999) have suggested
that there could be sectors of faster wind driven by luminous magnetic
rotator forces.

In the optical, a correlation has been found by Henrichs \etal\ (1998)
between the \Halpha\ emission line variability and the wind UV lines. This
links the DAC phenomena, discussed above, to something occurring at or near
the surface of the stars. The \Halpha\ double peaked lines are the best
studied diagnostic of the Be star disks. In recent years, monitoring of the
variations of the violet (V) and red (R) peaks in the \Halpha\ lines has
shown that there is a quasi cyclical change in the V and R components of the
line on time scales of a few years, (Hanuschik \etal\ 1995). This has been
explained with models for a one-armed spiral pattern or a ``global disk
oscillation'' in the equatorial disk around the stars (Okazaki, 1991, 1997;
Papaloizou \etal\ 1992) . For such a pattern to develop, the matter must
have a long residence time in the disk. This would not expected in the
``wind compressed disk'' (WCD) model of Be stars by Bjorkman and Cassinelli
(1993) that led to explanations of large equatorial density enhancements and
to the IR excess and intrinsic polarization of the Be stars. In the WCD
model matter from the star flows towards, and shocks at, the equatorial
region as a natural consequence of equator-crossing trajectories of a wind
flowing from a rapidly rotating star. However, a WCD has matter continuing
to flow outwards in the equatorial plane, and would not produce the one
armed spiral behavior. To achieve the large angular velocities of a
Keplerian disk, the star must transmit not only mass but also angular
momentum to its wind. This can be achieved by having the flow channeled by a
co-rotating magnetic field.

\section{The General Dynamical Effects of B fields on Hot star Winds}

The fast rotation of hot stars leads to the most prominent effects that
connect fields with the hot star outflows. Here several aspects are
considered. A topic which I call Luminous Magnetic Rotator theory was
developed by Friend \& MacGregor (1984), and applied to Wolf Rayet stars by
Poe, Friend \& Cassinelli (1987). It is a combination of the well know Weber
and Davis (1967) model for the solar wind, with line driven wind theory of
CAK. From the perspective of the basic Weber and Davis model, there is a
primary mechanism for driving an outflow, and it is amplified by the
effects of the magnetic field. The primary mechanism for the solar wind
would be combination of effects that could be called the the coronal
mechanism, for hot stars the primary mechanism would be the CAK wind theory.
The addition of a field can lead to the star being a slow, fast, or
centrifugal, magnetic rotator. The Sun is an example of a slow magnetic
rotator because the rotation of the field does not lead to a significant
increase in the wind speed. Fast Magnetic Rotator theory as explained
by Hartmann and Macgregor (1980) can lead to a fast equatorial outflow,
which could be important in the production of the co-rotating interaction
regions discussed above. In the centrifugal magnetic rotators the sub-sonic
region is rotating at roughly solid body rotation with the star and this
enhances the density at the critical point and leads to an increased mass
loss rate. Also along this sequence of magnetic rotators is an increase in
the angular momentum transfered from the star to the wind. It is the angular
momentum addition that makes it possible to explain the formation of disks
around stars.

An oblique rotator structure for the magnetic field is a particularly well
studied variant of the magnetic rotator theory. Shore (1999) has shown that
the B fields in oblique rotators give rise to a periodic and directed wind
that explains the outflows observed in B chemically peculiar stars. It is
this picture of a significant flow from the oblique rotating wind that is
supported by the time variations of the Zeeman effect that has been reported
by Henrichs \etal\ (2001). The theoretical picture for disk formation around
an oblique rotator has been developed for the Ap stars by Babel and
Montmerle (1997). It was used several times during this meeting to 
explain the observed cyclical variability of Ap stars.

In the formation of planetary nebulae, the magnetic fields have been used
to explain the bipolar outflows that are commonly seen. These models are
commonly called Magnetic Wind Blown Bubbles and the theory has been developed
by Chevalier and Luo (1994), Garc\'{i}a-Segura \etal\ (1999) and others.

\section{Timescales}

In hearing the talks on early- type stars at this meeting, it occurred to me
that there are some interesting stellar aging effects in hot stars that might be
explainable in a straightforward way.

We have heard from Mathys that among the A-stars, the Ap phenomenon does not
occur for the first 30 percent of the lives of the stars. Similarly, at the
recent IAU Coll. 175 meeting on Be stars, Capilla, Fabregat \& Baines,
(2000) and others reported that Be stars are not found in clusters that are
younger than about $10^7$ years, i.e. there is a delay time of about 10 - 30
percent of the life of a B stars before disks become an important feature.
So both types of magnetically induced phenomena appear to have a delay of
order 10 percent of the stellar life before they can occur. Stretching
further to the O-stars we find from Wood \& Churchwell (1989) that the stars
are hidden from view for the first 10 percent of their main-sequence
lifetimes. The first evidence of the O- stars are Ultra Compact HII regions,
which often show very massive bi-polar outflow which seem to me to be caused
by magnetic rotator forces. Churchwell (1997) tabulates several such O-stars
in which the mass in the bipolar outflow exceeds the mass of the star at the
center. Only magnetic forces associated with the collapse of the cloud could
produce such a situation.

Thus, there are interesting magnetic phenomena occurring in
the early lives of O, B, and A stars. In the case of the B and A stars, we
can certainly ask what could cause the delay the start of the Be and Ap
phenomena.

The time scales involved are very similar to the times mentioned in the talk
by Charbonneau. He showed results from the calculations of MacGregor and
Cassinelli (2001) for the rise of magnetic fields through the sub-adiabatic
envelopes of hot stars. For the calculation, a torus of flux that is similar
to those used in modeling the rise of magnetic fields in the Sun is
considered. Assuming an initial pressure and temperature balance between the
flux tube and the external medium, there is an initial buoyant force that
causes the field region rise. However being in a sub-adiabatic region the
rising tube finds itself cooler than the surroundings and has a tendency to
stop its rise. It continues to rise at a terminal rate that is set by
heating of the tube by the somewhat hotter ambient medium. This leads to a
rise of the flux tubes on a time scale comparable to roughly 30\% of a
lifetime delay for the onset of the Ap and Be phenomena. It seems reasonable
to suggest that some sort of flux tube rise scenario could produce what
we see in these two classes of stars. (A somewhat similar scenario has been
proposed at this meeting for the cool star dividing line on the HR diagram,
by Holzwarth, Sch\"{u}ssler, and Solanki). If the rise time for the flux
tube controls the onset of the magnetic phenomena in Ap and Be stars, it
leads us to also predict a termination of the Ap and Be phenomena for the
following reason. The composition of the interface region where the field is
generated is changing with time as the core becomes more enriched in helium.
Thus the mean molecular weight changes by about a factor of two and this
extra mass per particle could inhibit the rise of the magnetic flux tubes.

\section{Conclusions}

As a closing remark, I want to say that I have found this to be an extremely
useful meeting, because it has brought together astronomers with such a wide
range of range of interests and expertise. We in the hot star community have
relatively few actual measurements of magnetic fields, so it has been
especially beneficial to discuss and hear about magnetic mechanisms, effects
and diagnostics that used in solar and cool star astronomy research. I hope
that the different range in phenomena that is found in hot star astronomy
will continue to attract the interest of experts in the these other fields
as well.

\pagebreak

\newpage

\section*{Discussion}
\noindent MOUSCHOVIAS: You stated that magnetic fields can reverse
the infall and convert it to an outflow. Is that statement a result of a
calculation or is it speculation.\\[.1cm]

\noindent CASSINELLI: No, it is not a model result. However I think the
idea that an infall can be reversed has already been developed. Such a
reversal is calculated to occur in the Shu's X-wind model for the formation
of low mass stars. I was just suggesting that a similar thing could occur
for massive stars but at a larger distance from the star because of the
greater velocities and mass fluxes are involved.  I made the statement
because I am aware of the rate at which mass can be driven from a star by
the various forces available. It would be impossible to have such a massive
flow arise from the O star star itself, even by the Luminous Magnetic
Rotator theory, which drive the most massive of the winds we considered in
Lamers and Cassinelli (1999). The flow reversal would need to occur before
the matter became too deep in the stars gravitational field, and centrifugal
magnetic rotator forces would seem like the most likely ones to achieve the
outflow.\\[.1cm]

\noindent BERDYGINA: Be stars appear also in binaries. What will be the
difference between the Be star accreting matter from the secondary and a Be
star with the magnetic field without accretion. \\[.1cm]

\noindent CASSINELLI: The answer to this question has been addressed by
Geis (2000). Yes there are some Be stars that are binaries, but it is
now generally agreed that what is considered to be the ``Be phenomenon''
is not a mass transfer process, but arises most frequently in single stars.
\\[.1cm]

\noindent SCHMITT: Assume that the X-ray emission of hot stars is from
confined magnetic loops. 1) What is heating the plasma? and 2) Why is L$_x$
$\sim$ L$_{bol}$.\\[.1cm]

\noindent CASSINELLI: Even early-type stars can
have sources of mechanical energy, The stars tend to be rapid rotators and
the rotation is not solid body, but differential. In addition there are
circulation currents occurring in the stars and non-radial pulsations. Also,
in one well studied star, $\tau$ Sco B0.5 V, there seems to be an infall of
matter from the stalled wind. I certainly don't know of the mechanism that
could lead to heating, but there is no lack of possibilities for sources of
mechanical energy. The proportionality between L$_x$ and L$_{bol}$ holds
approximately for the O- stars, but these have winds that are optically
thick to X-rays. Most of the X-rays we detect are produced by shocks in the
winds, which are produced by interactions of matter driven by the luminosity
of the star. When we looked to the somewhat less luminous near main sequence
stars in our ROSAT studies, which have optically thin winds, and so all of
the X-rays that are produced can be seen (Cohen \etal 1997), we found that
L$_x$ dropped sharply below the $10^{-7}$ L$_{bol}$ relation, i.e. L$_x$ is
not proportional to L$_{bol}$.\\[.1cm]

\noindent MOSS: The rise speed of the flux tube is a stably stratified region 
is proportional to $B^2/r_T^2$, where $r_T$ is the tube radius and B the field
strength.  I estimate (see my paper) that for $t_{rise} \sim 10^8$ years, $r_T
\sim 10^{-3}$ R$_*$ for Ap star parameters. Thus the field rises as
``spaghetti'', but must become organized it the surface to present the large
scale (ordered) poloidal field. Can you comment on how this might 
occur?\\[.1cm]

\noindent CASSINELLI:  In the simple picture of flux tubes that MacGregor
and I have worked on, and as described in Charbonneau's talk, it is
difficult to picture how there could be the communication between the
northern and southern hemispheres of the star needed for the global poloidal
structure. Perhaps the rising flux tubes are primarily contributing to the
quadrupolar and other higher order components that we have learned here are
on present on Ap stars.\\[.1cm]

\noindent RUEDIGER: If the Ap stars are an old population among the A stars,
also their slow rotation must suddenly appear during the main-sequence
evolution. Or is the slow rotation of the Ap stars already present at the
ZAMS, without stellar magnetism and peculiar chemistry?\\[.1cm]

\noindent CASSINELLI: I wouldn't say that the delay of 10 to 30 %
would make the Ap stars an old population. But more to the point I do not
see that the slow rotation and the chemical peculiarity have to appear
simultaneously during the main sequence evolution. The diffusion of the
elements in Ap stars could be affected either by the presence of a dipole or
by higher the more localized fields associated with the rise of flux tubes.
\end{document}